\documentstyle[epsfig,12pt]{article}

\begin{document}
\begin{titlepage}
\begin{flushright}
UFIFT-HEP-02-25 \\ hep-th/0207191
\end{flushright}
\vspace{.4cm}
\begin{center}
\textbf{The Ostrogradskian Instability of \\
Lagrangians with Nonlocality of Finite Extent}
\end{center}
\begin{center}
R. P. Woodard$^{\dagger}$
\end{center}
\begin{center}
\textit{Department of Physics \\ University of Florida \\
Gainesville, FL 32611 USA}
\end{center}

\begin{center}
ABSTRACT
\end{center}
I reply to the objections recently raised by J. Llosa to my constructive
proof that Lagrangians with nonlocality of finite extent inherit the full
Ostrogradskian instability.

\begin{flushleft}
\end{flushleft}
\vspace{.4cm}
\begin{flushleft}
$^{\dagger}$ e-mail: woodard@phys.ufl.edu
\end{flushleft}
\end{titlepage}

\section{Introduction}

Consider a nonlocal Lagrangian $L[q](t)$ whose dynamical variable is $q(s)$.
I define the nonlocality to be of finite extent $\Delta t$ if $L[q](t)$
definitely depends upon (and mixes) $q(t)$ and $q(t + {\Delta t})$, and 
potentially depends as well upon $q(t')$ for $t < t' < t + {\Delta t}$. Two
years ago I constructed a Hamiltonian formalism for such systems \cite{RPW}.
(One should note the very similar earlier work by Llosa and Vives \cite{LV}.)
I also showed that this Hamiltonian could be obtained as the infinite 
derivative limit of the well-known construction of Ostrogradski \cite{Ostro} 
for nondegenerate higher derivative Lagrangians. 

Ostrogradski's construction \cite{Ostro} exhibits a virulent instability
which grows worse as the number of higher derivatives increases. If the 
Lagrangian depends nondegenerately upon $N$ time derivatives then there must 
be $N$ canonical coordinates and $N$ conjugate momenta, and the Hamiltonian 
is {\it linear} in $N-1$ of the momenta. Such a Hamiltonian can never be 
bounded below. This is what I mean by the ``Ostrogradskian instability''. The 
result is completely nonperturbative. Further, it should not be altered by 
quantum effects because the problem derives from a huge volume of the classical
phase space. The fact that Lagrangians with nonlocality of finite extent can 
be viewed as the infinite derivative limit of higher derivative Lagrangians
means that one can forget about any phenomenological applications for them. I 
illustrated the problem by explicitly solving a simple free theory of this 
type and demonstrating the existence of an infinite number of runaway 
solutions.

A recent comment by Llosa \cite{Llosa} disputes my negative assessment. His 
three criticisms are laid out in sections II, III and IV of his comment. 
Because of the continuing interest in nonlocal Lagrangians 
\cite{GKL,JM,B,M,GKM,CHY,CLZ,MZ} these replies to his comments may be of 
interest. This work does not pretend to be self-contained. One should read it 
in the context of my paper \cite{RPW} and Llosa's comment \cite{Llosa}. Each 
of my sections deals with the corresponding section of Llosa's comment. 
Because the equation numbers have changed between his first draft and the 
version subsequently accepted for publication I will parenthesize the new 
numbers and bracket the old ones.

\section{Reply to the comments of Section II}

Llosa's first objection mentions ``defects . . . concerning the 
application of the variational principle''. This is a curious criticism 
because my paper did not state, nor did it apply any variational principle. I
began with the nonlocal Euler-Lagrange equation and worked from there.

Of course there is a variational principle behind my equation of motion. It 
is not based on the action given in Llosa's equation (4) \{4\} but rather 
upon the form which comes from integrating the Lagrangian between arbitrary 
fixed times as usual,
\begin{equation}
S[q] = \int_{t_1}^{t_2} ds L[q](s) \; .
\end{equation}
The equation of motion is the variational derivative of the action at any time 
$t$ outside the influence of the initial and final surfaces, that is,
\begin{equation}
t_1 + {\Delta t} < t < t_2 - {\Delta t} \; . \label{range}
\end{equation}
Consider the nonlocal harmonic oscillator discussed in my paper,
\begin{equation}
L[q](t) = \frac12 m \dot{q}^2(t + {\Delta t}/2) - \frac12 m \omega^2
q(t) q(t + {\Delta t}) \; . \label{HO}
\end{equation}
For $t$ in the range (\ref{range}) one does not encounter surface terms 
from partial integration and the action's functional derivative gives,
\begin{eqnarray}
{\delta S[q] \over \delta q(t)} & = & m \int_{t_1}^{t_2} ds \left\{ \dot{q}(s
+ {\Delta t}/2) \delta'(s + {\Delta t}/2 - t) - \frac12 \omega^2 \delta(s - t)
q(s + {\Delta t}) \right. \nonumber \\
& & \hspace{4cm} \left. - \frac12 \omega^2 q(s) \delta(s + {\Delta t} - t)
\right\} \; , \\
& = & - m \left\{ \ddot{q}(t) + \frac12 \omega^2 q(t + {\Delta t}) + \frac12
\omega^2 q(t - {\Delta t}) \right\} \; .
\end{eqnarray}

The general result is the right hand side of equation (2) in my paper 
\cite{RPW}. It derives from the fact that an arbitrary Lagrangian $L[q](s)$
with nonlocality of finite extent ${\Delta t}$ is defined to depend upon 
$q(t)$ only for $s \leq t \leq s + {\Delta t}$. The result is therefore 
unchanged if we restrict $s$ to the range $t - {\Delta t} \leq s \leq t$. Now 
make the change of variables $r = t - s$,
\begin{equation}
{\delta S[q] \over \delta q(t)} = \int_{t_1}^{t_2} ds {\delta L[q](s) \over
\delta q(t)} = \int_0^{\Delta t} dr {\delta L[q](t-r) \over \delta q(t)} = 0
\; .
\end{equation}

In the original version of his comment Llosa maintained that the equation 
of motion should be defined as the full variational derivative of the action,
including surface terms. He has since altered this to conform to the usual
condition, for first order Lagrangians, of demanding that the variation vanish 
at the initial and final times. However, it is instructive to see what goes 
wrong when the original condition is applied. Consider the action integral for
the simple harmonic oscillator,
\begin{equation}
S[q] = \frac12 m \int_{t_1}^{t_2} ds \left\{ \dot{q}^2(s) - \omega^2 q^2(s) 
\right\} \; .
\end{equation}
Its full variational derivative is,
\begin{equation}
{\delta S[q] \over \delta q(t)} = -m \left\{ \ddot{q}(t) + \omega^2 q(t)
\right\} + m \dot{q}(t_2) \delta(t_2 - t) - m \dot{q}(t_1) \delta(t_1 - t)
= 0 \; . \label{firstorder}
\end{equation}
The surface variations enforce the conditions $\dot{q}(t_1) = 0= \dot{q}(t_2)$.
Unless $\omega (t_2 - t_1)$ happens to be an integer multiple of $\pi$ the
only solution is $q(t) = 0$. 

Most physicists regard the harmonic oscillator's solution space as two 
dimensional, characterized by the initial position and velocity. If one instead
demands the full variational derivative to vanish then the solution space is 
either zero or one dimensional, depending upon whether or not $\omega (t_2 - 
t_1)$ is an integer multiple of $\pi$. The distinction between zero and one 
does not derive from any property of the system's dynamics but rather from 
the completely arbitrary choice of the range of temporal integration in what 
we call the action. Further, it is difficult to reconcile {\it either} 
dimensionality with one's experience --- with swings or masses on springs --- 
that an oscillator can be released with a very wide range of initial positions 
and initial velocities. Hence we conclude that action principles should be 
formulated so as to determine evolution from an arbitrary initial condition, 
not to fix initial and final conditions. 

The standard procedure for Lagrangians involving the variable and its first
derivative is to demand stationarity with respect to variations which vanish
at the initial and final point. This is not done because there is anything
special about the initial and final times, or about the zeroth derivative of
the variation, but rather because it suffices to null the surface terms coming 
from partial integration. Achieving the same end --- the absence of surface 
contributions --- requires more restricted variations as one adds higher 
derivatives. For each higher derivative in the Lagrangian one more derivative 
of the variation is required to vanish. Extending these considerations leads
one to demand stationarity with respect to variations which vanish outside 
range (\ref{range}). Thus my action principle is not ``nonstandard'' but 
rather a straightforward generalization of the usual procedure for freeing 
what one calls, ``the Euler-Lagrange equations'' from surface terms. 

Of course one can always attempt to constrain an unstable theory so as to 
achieve a stable one, and Llosa's initial and final conditions can be viewed
in this light. The big problem then is to make the constraints consistent 
with time evolution. This is trivial for theories with linear equations of 
motion but it is very difficult when nonlinearities are present, as they must 
be present for any sector of the universe which we can observe. In field 
theory one must also check that the constraints preserve causality and 
Poincar\'e invariance. It is so difficult to reconcile these three conditions 
for interacting field theories which suffer the Ostrogradskian instability 
that no one has ever succeeded in doing it. In any case I do not assert that 
no such constraints exist, only that the unconstrained theory is unstable.

Finally, one should note that the final argument in the revised version of 
Llosa's Section II is incorrect. The argument purports to show that there is 
a potential problem of consistency in specifying $q(t)$ for the ranges $t_1 
\leq t < t_1 + {\Delta t}$ and $t_2 - {\Delta t} < t \leq t_2$. The argument 
consists of using the equation of motion (without surface terms) to determine 
$q(t)$ in the latter region in terms of $q(t)$ in the former region. This is 
false. The equations of motion actually determine $q(t)$ in terms of its 
values in the region $t_1 - {\Delta t} \leq t < t_1 + {\Delta t}$, not $t_1 
\leq t < t_1 + {\Delta t}$. Hence the range of times is $2 {\Delta t}$, with 
either initial and final conditions or purely initial conditions. In this
regard the nonlocal system is analogous to that of a first order Lagrangian
for which one can specify either $q(t_1)$ and $q(t_2)$ or else $q(t_1)$ and 
$\dot{q}(t_1)$. Of course the fully initial value formulation is vastly
preferable, both because it avoids the possibility of ambiguous situations
such as $\omega (t_2 - t_1) = \pi N$ in the harmonic oscillator, and because 
it corresponds to our sense of being able to prepare systems and then observe 
their time evolution. This is one more reason for not including surface
variations in the action principle.

\section{Reply to the comments of Section III}

Section III purports to demonstrate that the nonlocalized harmonic oscillator
(\ref{HO}) is actually stable. The problem with Llosa's argument is
that it assumes the conclusion. The general solution consists, as he states,
of a superposition of the form,
\begin{equation}
q(t) = \sum_{\ell} \left(A_{\ell} e^{i k_{\ell} t} + A^*_{\ell} e^{-i k_{\ell}
t}\right) \; ,
\end{equation}
where the $k_{\ell}$'s obey the equation,
\begin{equation}
k^2 = \omega^2 \cos(k {\Delta t}) \; .
\end{equation}
As was proved in my paper \cite{RPW}, there is a countably infinite class of 
conjugate pairs of solutions with nonzero imaginary parts. This proves the 
existence of exponentially growing modes, which are a manifestation (although 
not an essential one) of instability. Llosa discards these solutions by 
imposing the condition that $q(t)$ and its first two derivatives are bounded 
functions of $t$. (By the way, why not more derivatives, or fewer?) It is not 
valid to check stability by first ruling out the existence of exponentially 
growing modes.

Suppose this same restriction were applied to the Lagrangian of a point 
particle moving on a concave parabolic hill of infinite extent,
\begin{equation}
L = \frac12 m \dot{q}^2 + \frac12 m \omega^2 q^2(t) \; .
\end{equation}
The equation of motion is,
\begin{equation}
\ddot{q}(t) - \omega^2 q(t) = 0
\end{equation}
and the general initial value solution is,
\begin{equation}
q(t) = q_0 \cosh(\omega t) + \frac1{\omega} \dot{q}_0 \sinh(\omega t) \; .
\end{equation}
Most physicists regard this system as unstable, and the exponentially growing
solutions as proof of the instability. But Llosa's criterion of 
boundedness would require $q_0 = 0 = \dot{q}_0$, and his argument could be
invoked to pronounce the system ``stable''.

Another problem with Section III, which recurs in Section IV.A, is Llosa's
definition of stability. The Ostrogradskian instability is that the energy
functional is unbounded below over essentially half of the classical phase
space. Again, I believe most physicists would regard this as evidence of
instability. Llosa instead defines stability as the boundedness of 
deviations under variation of the initial conditions. For a conventionally 
interacting field theory the two definitions typically agree. However, when 
interactions are turned off it is possible to have negative energy modes 
which do not violate Llosa's criterion. 

The simplest example is obtained by reversing the sign of the harmonic 
oscillator Lagrangian,
\begin{equation}
L = -\frac12 m \dot{q}^2 + \frac12 m \omega^2 q^2 \; .
\end{equation}
Of course this does not alter the general initial value solution,
\begin{equation}
q(t) = q_0 \cos(\omega t) + \frac1{\omega} \dot{q}_0 \sin(\omega t) \; ,
\end{equation}
so this system is ``stable'' in Llosa's sense. It is also completely
unobservable because it fails to interact with anything else in the universe. 
Suppose we couple it to electromagnetism. What happens is obvious: the 
oscillator jumps down to its first excited state, and the energy difference 
comes off as a photon. This process is overwhelmingly likely to go because 
there is only one state with the oscillator unexcited and no photons, versus 
an infinite number of directions for the photon to take. Nor do things stop 
there. It is just as favorable for the oscillator to jump to its second excited 
state and emit another photon. In fact the oscillator will continue jumping 
to lower and lower energies, without bound. That is why it makes sense to use 
the energy definition of instability, and why the application of Llosa's 
criterion to a noninteracting system gives a misleading answer.

\section{Reply to the comments of Section IV}

Llosa's Lagrangian (25) \{16\}, given in Section IV.B, is an example of 
nonlocality of finite extent. However, the associated equation of motion is 
not Llosa's equation (28) \{19\} but rather,
\begin{eqnarray}
\lefteqn{\int_0^T dr {\delta L[q](t-r) \over \delta q(t)} = -\ddot{q}(t) - 
\omega^2 q(t) + \frac12 \omega^4 \int_0^T dt' G(t,t') q(t') } \nonumber \\
& & \hspace{4cm} + \frac12 \omega^4 \int_{t-T}^t ds G(s,t) q(s) = 0 \; .
\end{eqnarray}
Note that acting $(d/dt)^2 + \omega^2$ does not result in a local equation on
account of the final term, which is absent in Llosa's equation of motion. Hence
Llosa's further comments concerning his equation of motion are not relevant 
to the stability of systems with nonlocality of finite extent.

Although Llosa's equation (28) \{19\} is not the variation (in the standard 
sense) of his Lagrangian (25) \{16\}, it does have an interesting 
interpretation. It is what comes from integrating out the variable $x(t)$ in 
the following local, first derivative system,
\begin{eqnarray}
\ddot{q}(t) + \omega^2 q(t) & = & \omega^2 x(t) \; , \\
\ddot{x}(t) + \omega^2 x(t) & = & \omega^2 q(t) \; ,
\end{eqnarray}
subject to the conditions $x(0) = 0 = x(T)$. It is these two conditions which 
result in the original 4-dimensional space of solutions degenerating to the 
2-dimensional space that Llosa finds. Of course there is no problem with the 
stability of this system, nor does my paper assert otherwise. In fact the 
first page of my paper states, {\it ... the higher derivative representation 
is certainly not valid for the inverse differential operators which result 
from integrating out a local field variable.}

\section{Discussion}

Most of Llosa's comments have little to do with my work on nonlocality. What
they challenge instead is the view that the Ostrogradskian instability is a 
problem for nondegenerate higher derivative Lagrangians. The propensity to
raise such a challenge deserves comment in its own right.

It has long seemed to me that the Ostrogradskian instability is the most 
powerful, and the least recognized, fundamental restriction upon Lagrangian 
field theory. It rules out far more candidate Lagrangians than any symmetry 
principle. Theoretical physicists dislike being told they cannot do something
and such a bald no-go theorem provokes them to envisage tortuous evasions. I 
do not believe Llosa has discovered any way around the problem, but it is
impossible to demonstrate that none exists. I do urge the application of 
common sense. The Ostrogradskian instability should not seem surprising. It
explains why every single system we have so far observed seems to be described,
on the fundamental level, by a local Lagrangian containing no higher than first
time derivatives. The bizarre and incredible thing would be if this fact was
simply an accident.

\vskip 1cm
\centerline{\bf Acknowledgments}

This work was partially supported by DOE contract DE-FG02-97ER\-41029 and 
by the Institute for Fundamental Theory.

\end{document}